\documentclass[slac_one]{revtex4}
\usepackage{graphicx}
\usepackage{fancyhdr}
\begin{document}

\newcommand{\dedx}{\mbox{${\rm d}E/{\rm d}x$}}

\title{Large-angle hadron production cross-sections for the neutrino
factory} 

%

\author{I. Boyko (for the 
HARP--CDP group\footnote{The members of the 
HARP--CDP group are:
A.~Bolshakova, 
I.~Boyko, 
G.~Chelkov, 
D.~Dedovitch, 
A.~Elagin, 
M.~Gostkin,
S.~Grishin,
A.~Guskov, 
Z.~Kroumchtein, 
Yu.~Nefedov, 
K.~Nikolaev and
A.~Zhemchugov 
from the Joint Institute for Nuclear Research, 
Dubna, Russian Federation;
F.~Dydak and  
J.~Wotschack 
from CERN, Geneva, Switzerland; 
A.~De~Min 
from the Politecnico di Milano and INFN, 
Sezione di Milano-Bicocca, Milan, Italy; and
V.~Ammosov, 
V.~Gapienko, 
V.~Koreshev, 
A.~Semak, 
Yu.~Sviridov, 
E.~Usenko and  
V.~Zaets 
from the Institute of High Energy Physics, Protvino, 
Russian Federation.} 
)}
\affiliation{Joint Institute for Nuclear Research, Dubna,
Russian Federation}

\begin{abstract}
Precise measurements of neutrino oscillation parameters
and of neutrino--nucleon cross-sections require a good 
understanding of neutrino beams: flux as a function of
energy, transverse beam profile, and flavour composition. 
For this, hadron production spectra in proton--nucleus
collisions are essential.  
We report on double-differential inclusive large-angle cross-sections of 
the production of secondary protons and charged pions, in the 
interactions with a 5\% $\lambda_{\rm abs}$ 
thick stationary beryllium target, of proton and pion beams with
momentum from $\pm3$~GeV/{\it c} to $\pm15$~GeV/{\it c}. 
Our results show
cross-sections reported by the `HARP Collaboration' to be 
wrong by factors of up to two. 
\end{abstract}

\maketitle

\thispagestyle{fancy}


\section{INTRODUCTION} 

Precise cross-sections of secondary hadron production from the
interactions of protons and pions with nuclei are, {\it inter alia\/},
of importance for the understanding of the characteristics of
muons from the decay of pions that are produced by the proton
driver of a neutrino factory. Surprisingly, 
inclusive differential cross-sections of hadron production 
in the interactions of few GeV/{\it c} protons with nuclei are 
known only within a factor of two to three. Consequently, 
the HARP detector was designed to carry 
out a programme of systematic and precise measurements of 
hadron production by protons and pions with momenta from 
1.5 to 15~GeV/{\it c}. 

The detector combined a forward spectrometer with a 
large-angle spectrometer. The latter comprised a 
cylindrical Time Projection 
Chamber (TPC) around the target and an array of 
Resistive Plate Chambers (RPCs) that surrounded the 
TPC. The purpose of the TPC was track 
reconstruction and particle identification by \dedx . The 
purpose of the RPCs was to complement the 
particle identification by time of flight.

The HARP experiment was performed at the CERN Proton Synchrotron 
in 2001 and 2002 with a set of stationary targets ranging 
from hydrogen to lead, including beryllium.

We have measured~\cite{Beryllium1,Beryllium2} the inclusive 
cross-sections of the large-angle production (polar angle $\theta$ 
in the range from 20 to 125$^\circ$) 
of secondary protons and charged pions in 
the interactions with a 5\% $\lambda_{\rm abs}$ beryllium target
of protons and pions with beam momenta of $\pm3.0$,
$\pm5.0$, $-8.0$, $+8.9$, $\pm12.0$, and $\pm15.0$~GeV/{\it c}.  

\section{DETECTOR CHARACTERISTICS AND PERFORMANCE}

\begin{figure*}[h]
\begin{center}
\begin{tabular}{cc} 
\includegraphics[height=0.3\textwidth]{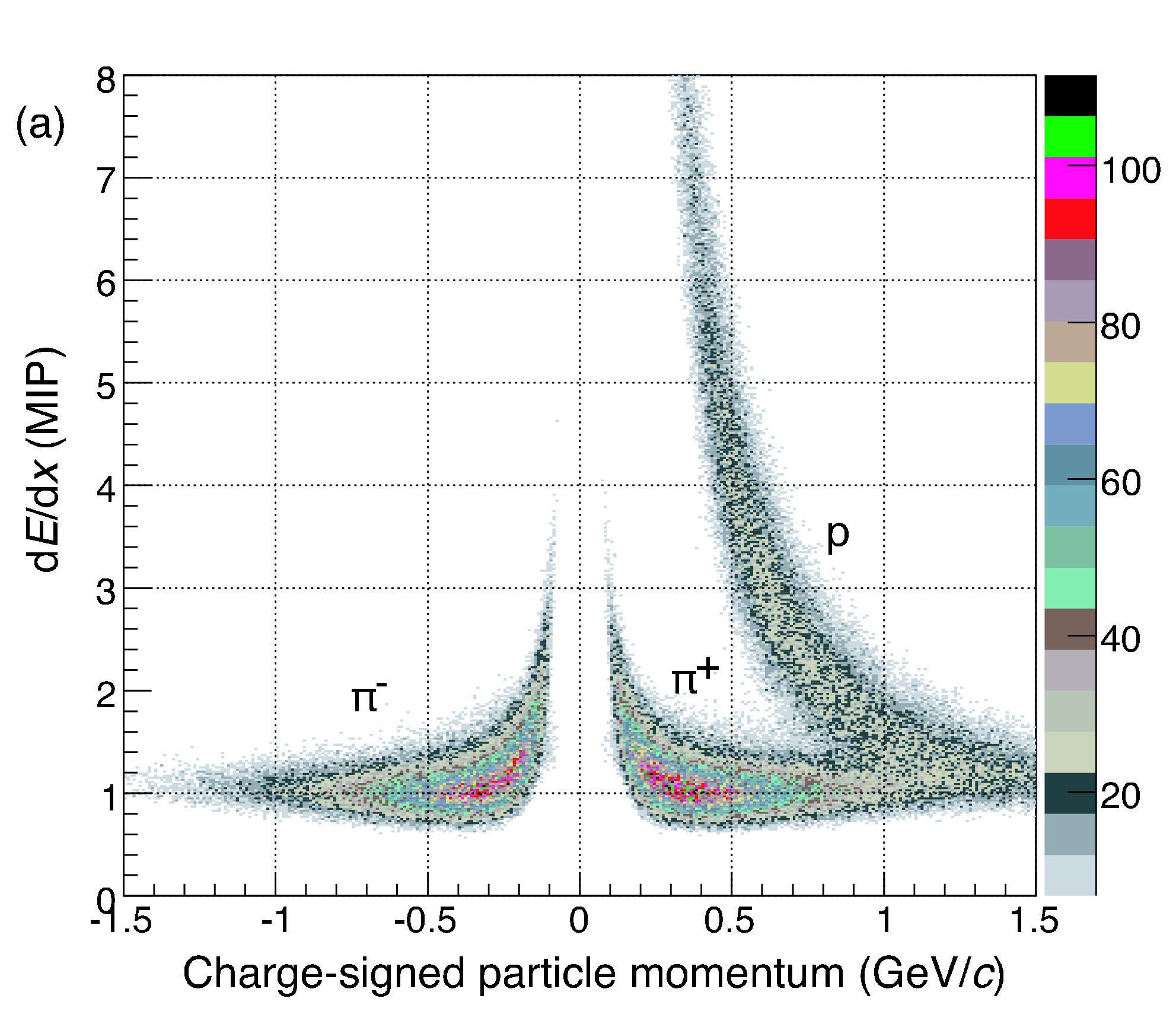} &
\includegraphics[height=0.3\textwidth]{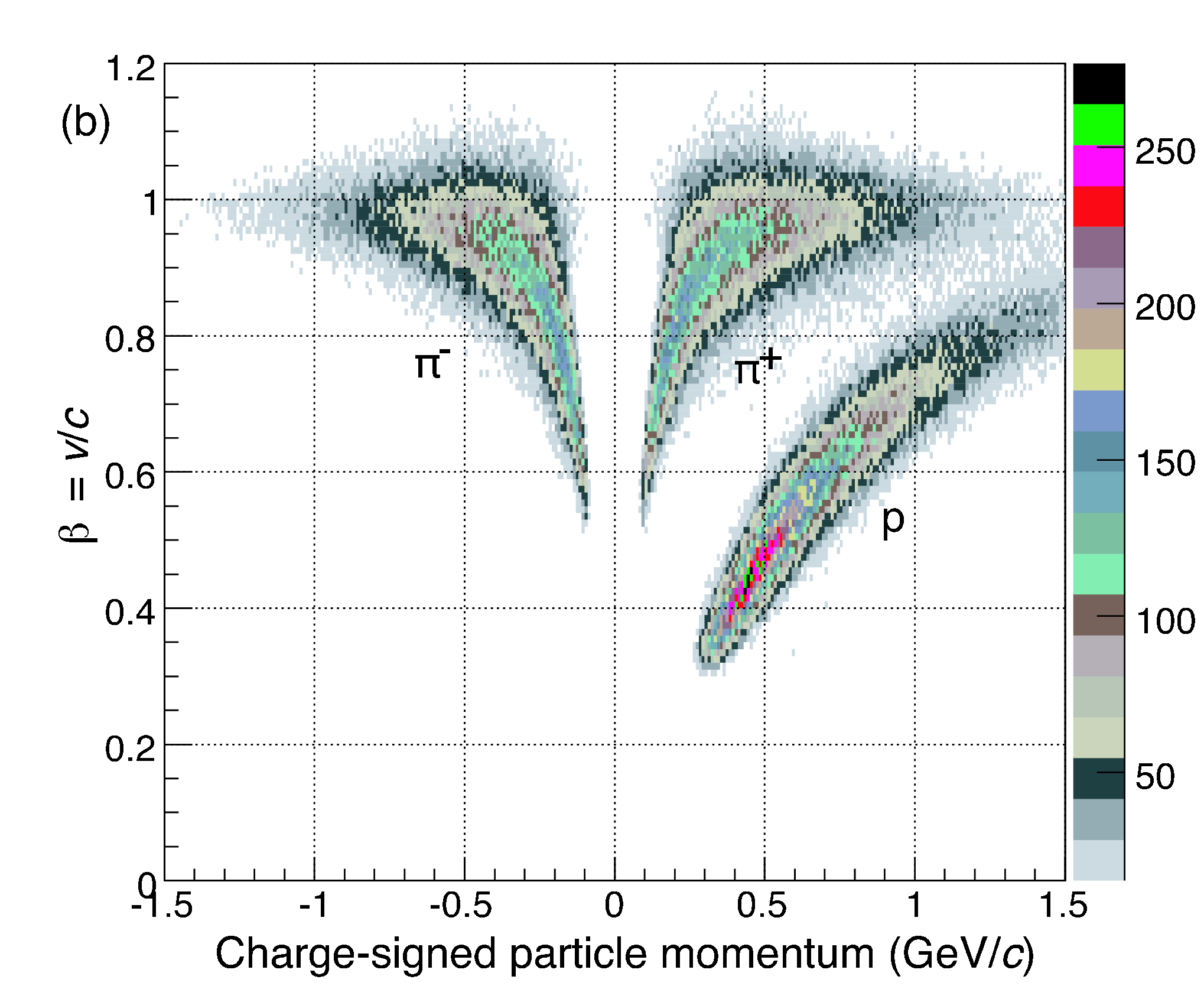} \\
\end{tabular}
\end{center}
\caption{Specific ionization d$E$/d$x$ (left panel) and velocity $\beta$
(right panel) versus the charge-signed momentum of positive and 
negative tracks 
in $+8.9$~GeV/{\it c} data.}
\label{dedxandbeta}
\end{figure*}

For the work reported here, only the HARP large-angle spectrometer
was used~\cite{TPCpub,RPCpub}. Its salient technical 
characteristics are stated in Table~\ref{LAcharacteristics}. The 
good particle identification 
capability stemming from \dedx\ in the TPC and from time of flight 
in the RPC's is demonstrated in Fig.~\ref{dedxandbeta}. 
\begin{table}[ht]
\vspace*{2mm}
\begin{center}
\caption{Technical characteristics of the HARP large-angle spectrometer}
\begin{tabular}{|c|c|}
\hline 
\textbf{TPC} & \textbf{RPCs}  \\
\hline
\hline
$\sigma(1/p_{\rm T}) \sim 0.20-0.25$~(GeV/{\it c})$^{-1}$ &
   Intrinsic efficiency $\sim 98$\%  \\
$\sigma(\theta) \sim 9$~mrad  & $\sigma$(TOF) $\sim 175$~ps  \\
$\sigma({\rm d}E/{\rm d}x) / {\rm d}E{\rm d}x \sim 0.16$  &  \\ 
\hline
\end{tabular}
\label{LAcharacteristics}
\end{center}
\end{table}

\section{INCLUSIVE CROSS-SECTIONS}

\begin{figure*}[ht]
\begin{center}
\begin{tabular}{ccc}
\includegraphics[height=0.25\textheight]{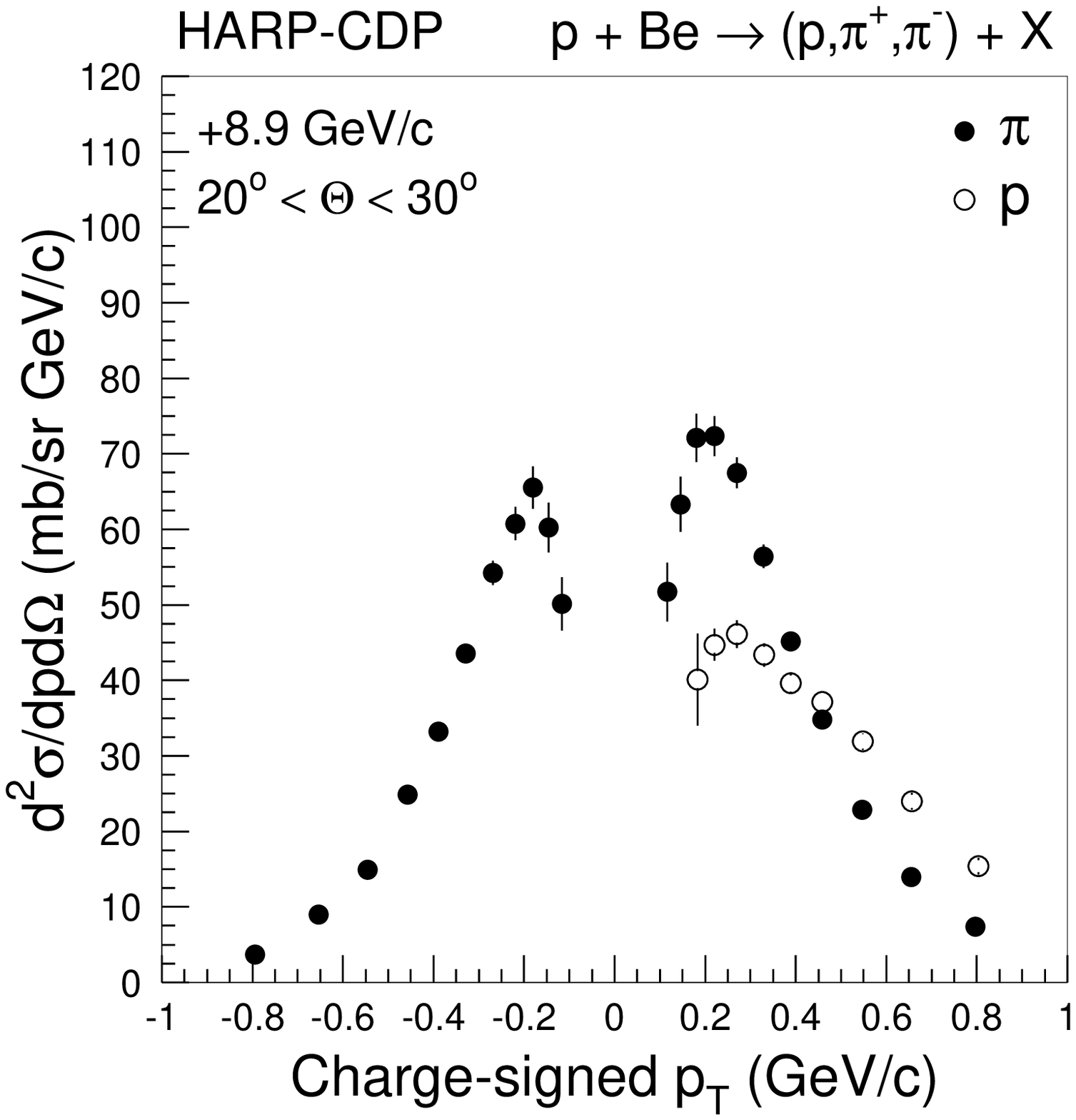} &
\includegraphics[height=0.25\textheight]{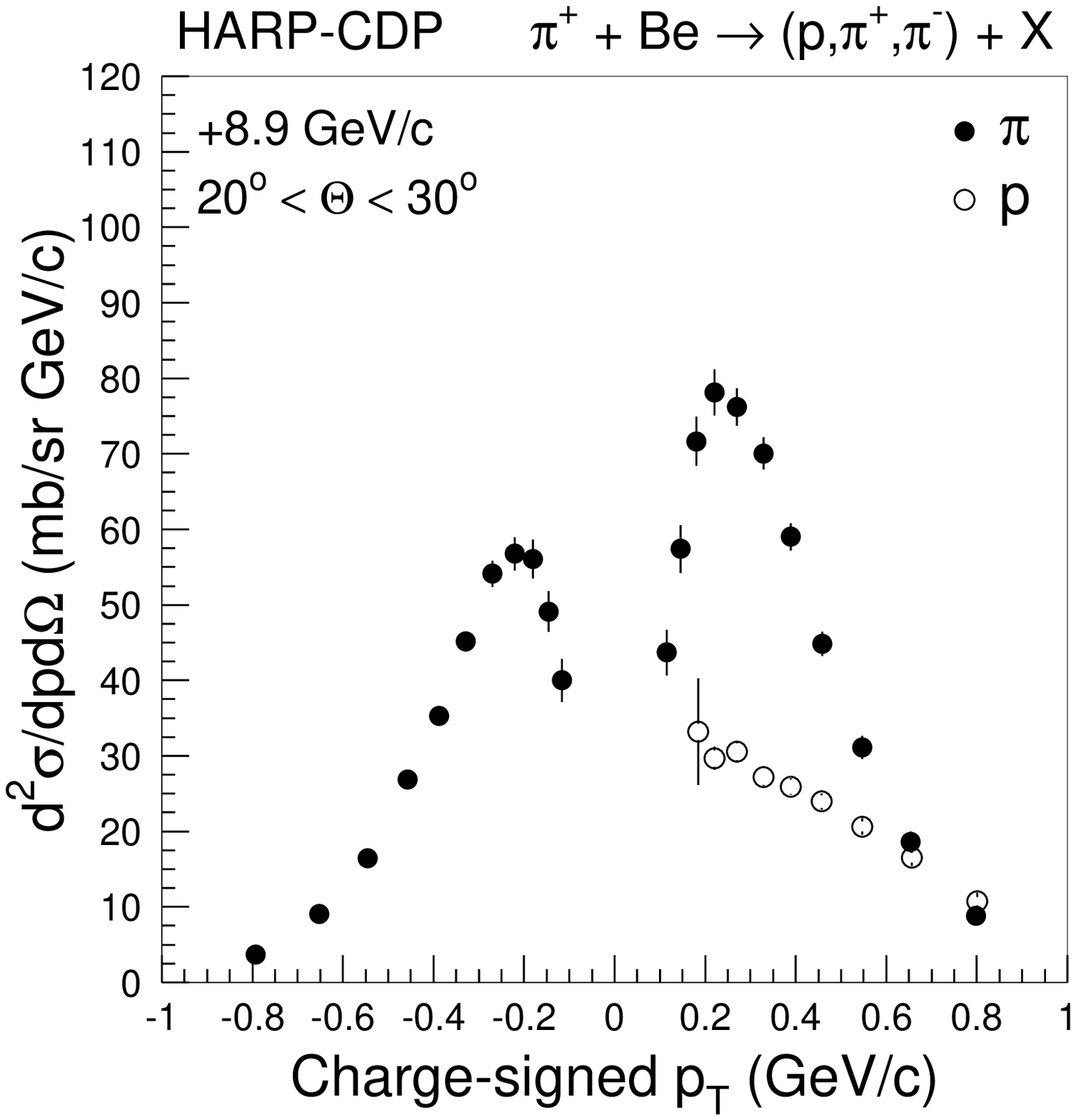} &
\includegraphics[height=0.25\textheight]{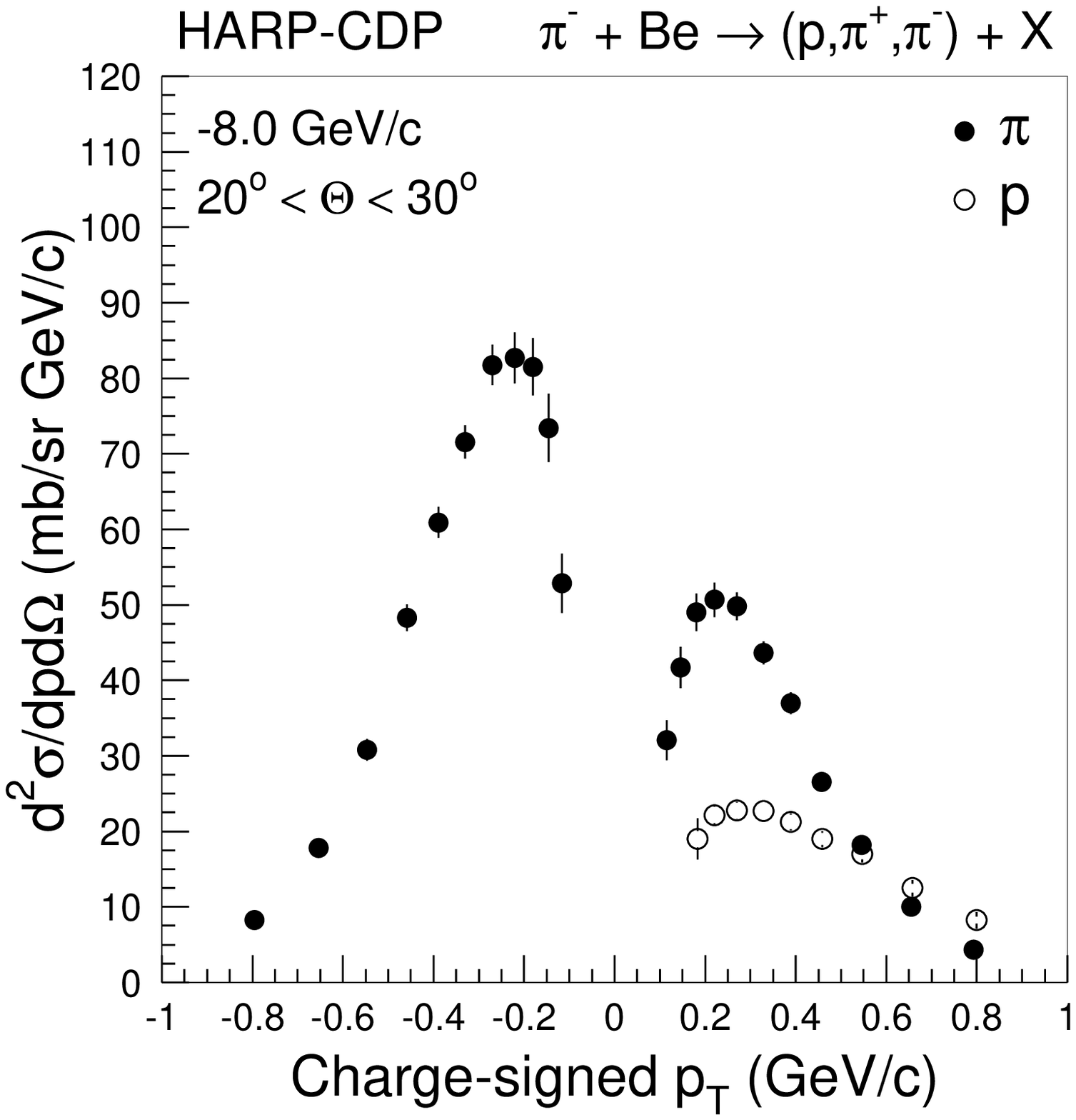} \\
\end{tabular}
\caption{Inclusive cross-sections of the production of secondary
protons, $\pi^+$'s, and $\pi^-$'s, in the polar-angle range 
20 to 30$^\circ$, by protons (left panel),
$\pi^+$ (middle panel) and $\pi^-$ (right panel) with 
beam momentum of $+8.9$~GeV/{\it c} on beryllium nuclei,  
as a function of the charge-signed 
$p_{\rm T}$ of the secondaries; the shown errors are total errors.} 
\label{xsvsmom}
\end{center}
\end{figure*}

Figure~\ref{xsvsmom} shows examples of inclusive cross-sections of
proton and $\pi^{\pm}$ production on beryllium nuclei. Double-differential
cross-sections with a typical precision of a few per cent are available
in tabular form in Refs.~\cite{Beryllium1,Beryllium2} for the polar-angle
range of 20 to 125$^\circ$ and for the $p_{\rm T}$ range 
of 0.10 to 1.25~GeV/{\it c}.

\section{COMPARISON WITH CROSS-SECTIONS FROM THE E802, E910, AND HARP
COLLABORATIONS}

\begin{figure*}[ht]
\begin{center}
\begin{tabular}{ccc}
\includegraphics[width=0.3\textwidth]{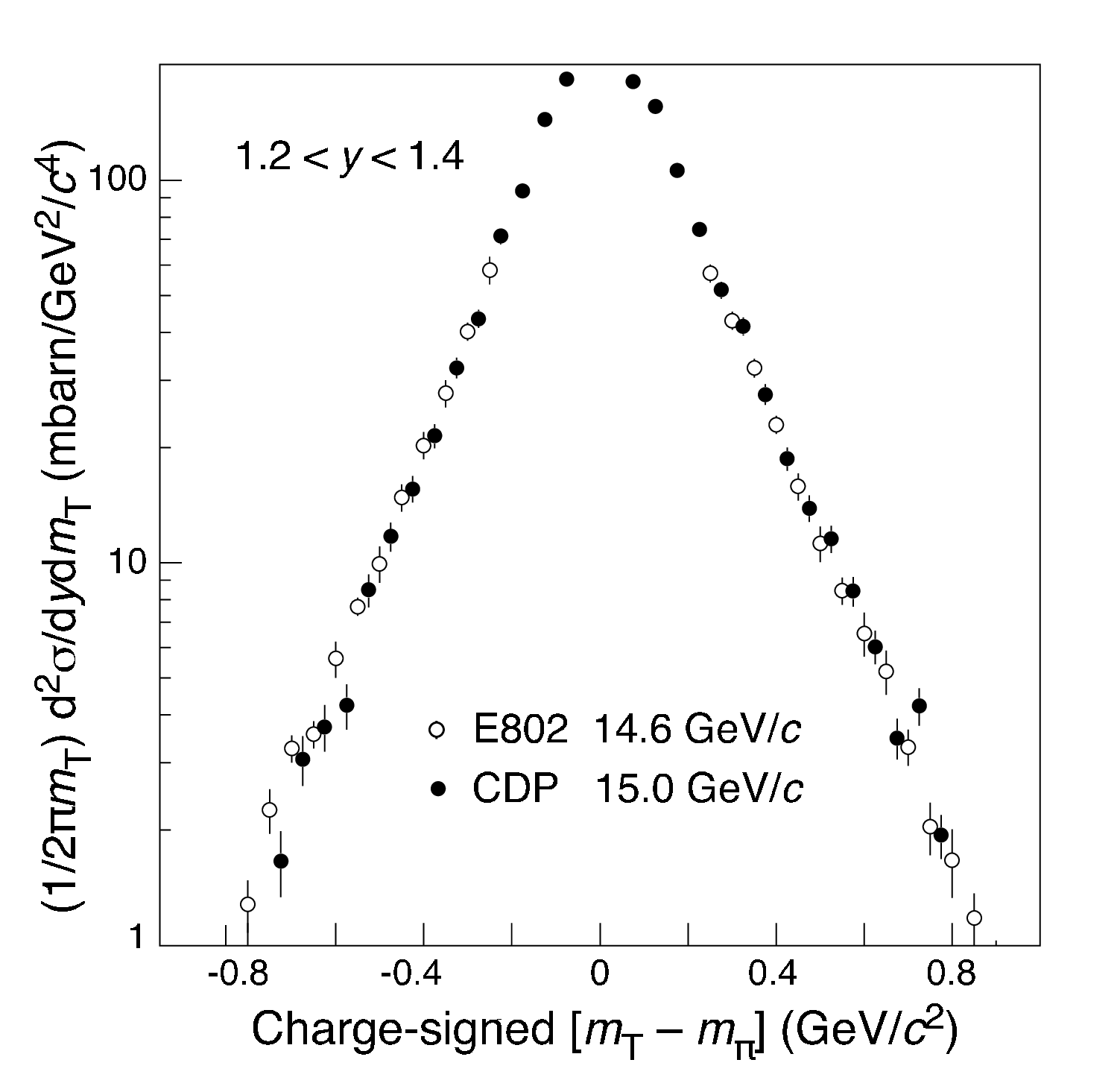} & 
\includegraphics[width=0.3\textwidth]{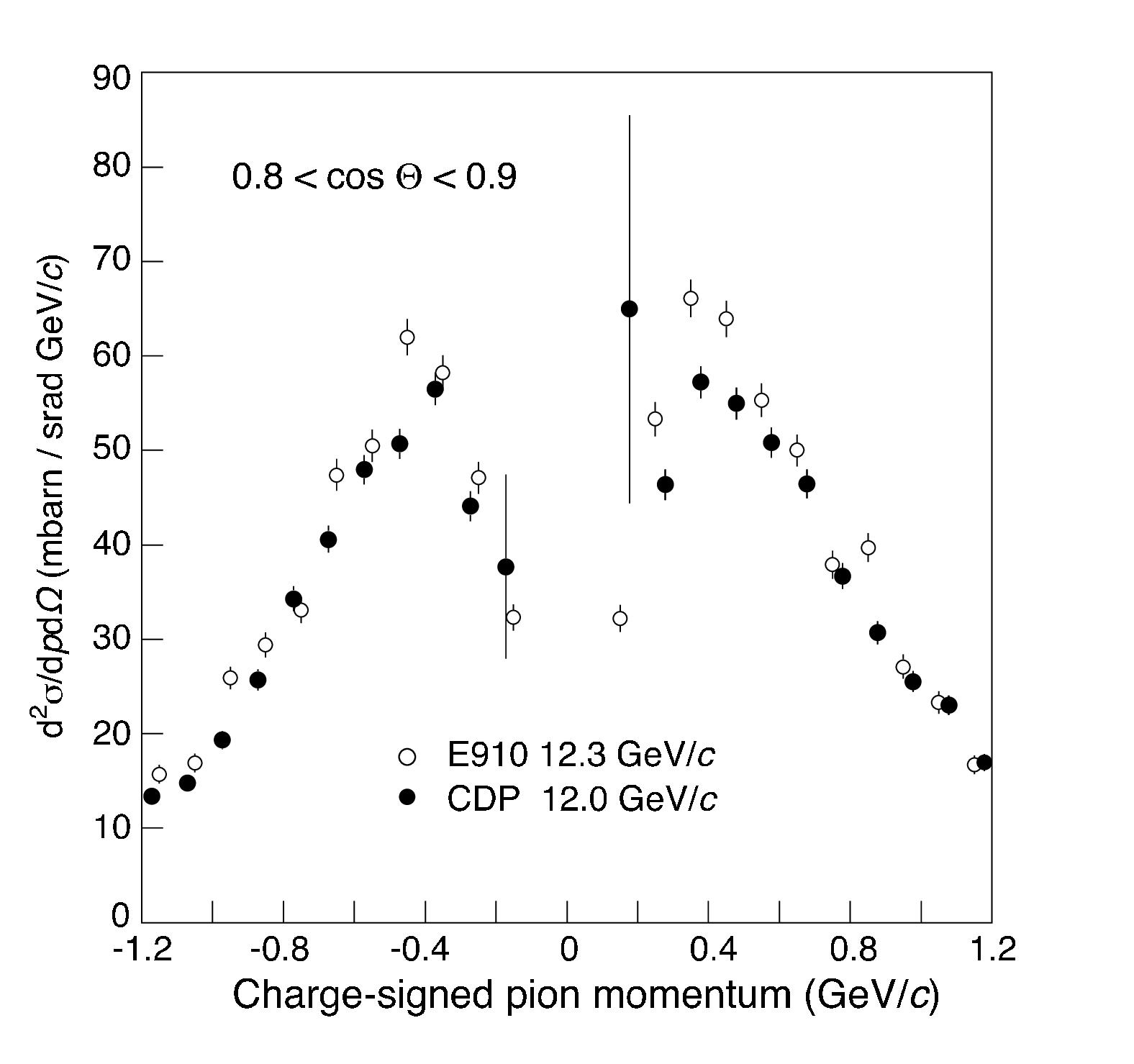} &
\includegraphics[width=0.3\textwidth]{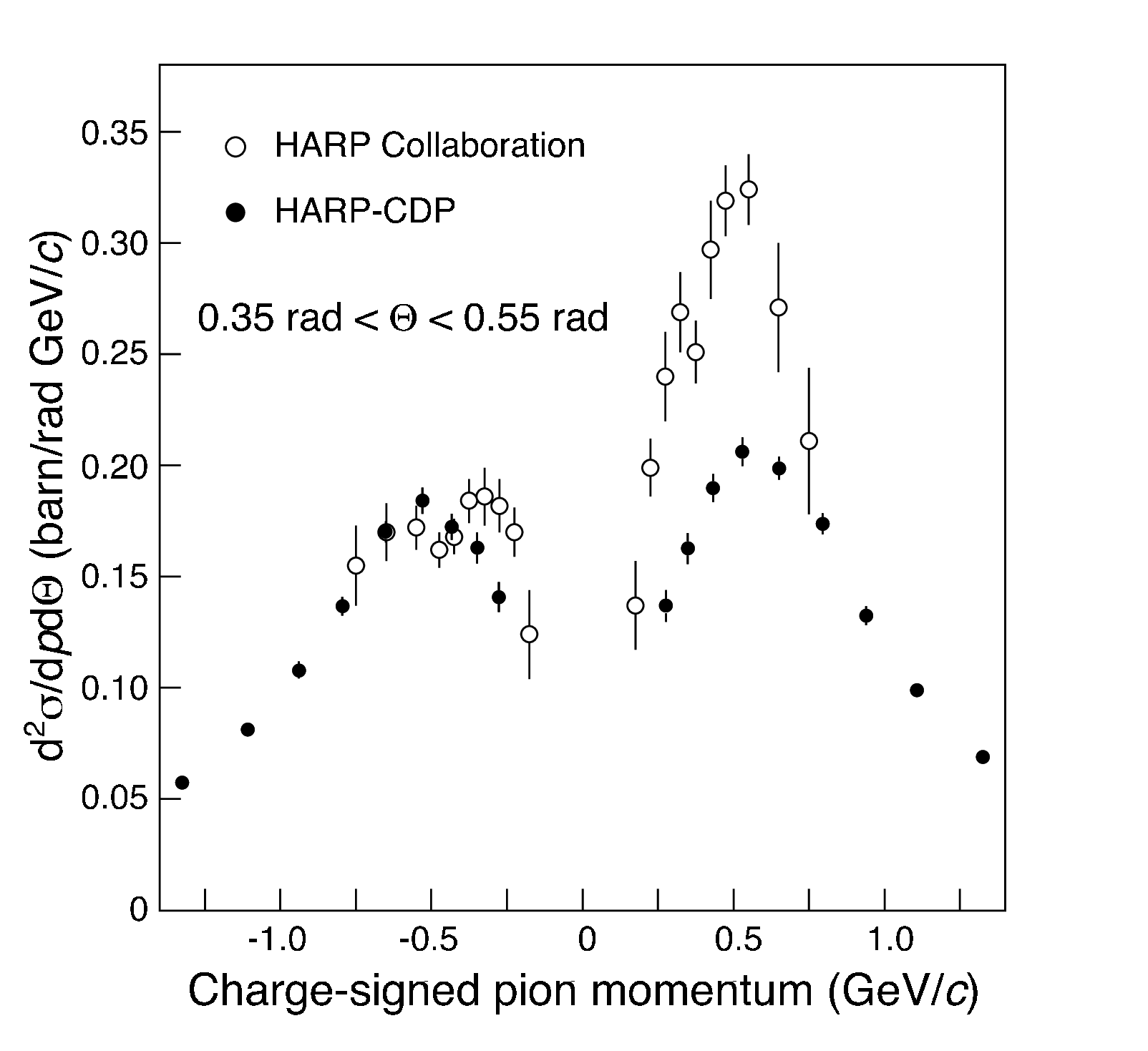} \\ 
\end{tabular}
\caption{Left panel: comparison of our cross-sections (black dots) 
of $\pi^\pm$ production by $+15.0$~GeV/{\it c} 
protons off beryllium nuclei with the cross-sections 
published by the E802 Collaboration for the proton beam 
momentum of $+14.6$~GeV/{\it c} (open circles); 
middle panel: comparison of our cross-sections (black dots)  
of $\pi^\pm$ production by $+12.0$~GeV/{\it c} 
protons off beryllium nuclei with the cross-sections 
published by the E910 Collaboration for the proton beam 
momentum of $+12.3$~GeV/{\it c} (open circles);
right panel: comparison of our cross-sections (black dots) 
of $\pi^\pm$ production by $+12.0$~GeV/{\it c} 
protons off beryllium nuclei with the cross-sections 
published by the HARP Collaboration (open circles).}
\label{comparisonwithE802andE910andOH}
\end{center}
\end{figure*}

The left panel in Fig.~\ref{comparisonwithE802andE910andOH} shows 
the Lorentz-invariant 
cross-section of $\pi^+$ and $\pi^-$ production by
$+14.6$~GeV/{\it c} proton interactions with beryllium nuclei, 
in the rapidity range $1.2 < y < 1.4$,
published by the E802 
Collaboration~\cite{E802}. Their data are compared 
with our cross-sections from the interactions of $+15.0$~GeV/{\it c} 
protons, expressed in E802 units. We note good agreement.

The middle panel in Fig.~\ref{comparisonwithE802andE910andOH} shows the
cross-section ${\rm d}^2 \sigma / {\rm d}p {\rm d}\Omega$ 
of $\pi^\pm$ production by $+12.3$~GeV/{\it c} protons,
in the polar-angle range $0.8 < \cos\theta < 0.9$, published by
the E910 Collaboration~\cite{E910}. Their data are compared 
with our cross-sections from the interactions of $+12.0$~GeV/{\it c} 
protons, expressed in E910 units. We note reasonable agreement.

The right panel of Fig.~\ref{comparisonwithE802andE910andOH} shows the 
HARP Collaboration's cross-sections~\cite{OffLApaper} 
of $\pi^\pm$ production by $+12.0$~GeV/{\it c} 
protons off beryllium nuclei. Their data are compared 
with our respective cross-sections, expressed in 
the units used by the HARP Collaboration. We note striking
disagreement.

As detailed in Ref.~\cite{JINSTpub} and in references cited 
therein, the HARP Collaboration's 
data analysis is affected by their 
lack of understanding of TPC track distortions which 
leads to (i) a bias of $\Delta (1/p_{\rm T}) 
\simeq 0.3$~(GeV/{\it c})$^{-1}$; (ii)
a resolution of $\sigma (1/p_{\rm T})  
\simeq 0.6$~(GeV/{\it c})$^{-1}$ which is by a factor of two worse
than claimed by them; and (iii) a bad overall RPC time-of-flight 
resolution of 305~ps and
an apparent advance of the timing signal of protons 
with respect to that of pions by $\sim$500~ps
(`500~ps effect'). All this causes distorted momentum spectra of secondary 
hadrons especially in regions where there is a 
strong momentum dependence, and the misidentification of 
protons as pions. The HARP Collaboration's pion 
production cross-sections are fatally biased and
unsuitable for the design of a neutrino factory.

\end{document}